\documentclass[12pt,preprint]{aastex}

\def\Msun{$M_{\odot}$}

\def\Lsun{$L_{\odot}$}

\def\kms{kms$^{-1}$}

\def\n2hhp{N$_{2}$H$^{+}$}

\def\h13cop{H$^{13}$CO$^+$}
\def\hc5n{HC$_{5}$N}

\def\t32{$J=3-2$}
\def\h13cn{H$^{13}$CN}
\def\cc34s{CC$^{34}$S}
\def\n2hp{N$_2$H$^+$}

\def\13co{$^{13}$CO}
\def\c18o{C$^{18}$O}
\def\ch3cn{CH$_{3}$CN}
\def\c34s{C$^{34}$S}
\def\3423{$3_4-2_3$}

\def\Tmb{$T_{\rm mb}$}
\def\Tex{$T_{\rm ex}$}

\def\deg{\hbox{$^{\circ}$}}
\def\arcmin{\hbox{$^{\prime}$}}
\def\arcsec{\hbox{$^{\prime\prime}$}}

\def\cc{cm$^{-3}$}

\def\kms{km s$^{-1}$}

\def\tmb{$T_{\rm mb}$}

\def\VFWHM{$V_{\rm FWHM}$}

\def\h2{$H_{2}$}
\def\reff350{$R_{\rm eff~350}$}

\slugcomment{Accepted for publication in the ApJL}

\shorttitle{Extended Warm Dense Gas At The Heart Of A Cold Collapsing Dense Core}
\shortauthors{Shinnaga et al.}

\begin{document}


\title{Warm Extended Dense Gas Lurking At The Heart Of A Cold Collapsing Dense Core}


\author{Hiroko Shinnaga\altaffilmark{1}, Thomas G. Phillips\altaffilmark{1,2}, Ray S. Furuya\altaffilmark{3}, and Yoshimi Kitamura\altaffilmark{4}
}



\altaffiltext{1}{California Institute of Technology Submillimeter Observatory (CSO)
 111 Nowelo St. Hilo HI 96720}
\altaffiltext{2}{Division of Physics, Mathematics, and Astronomy, California Institute of Technology 320-47, Pasadena CA 91125}
\altaffiltext{3}{Subaru Telescope, National Astronomical Observatory of Japan, 650 North A`ohoku Place, Hilo, HI 96720}
\altaffiltext{4}{Institute of Space and Astronautical Science, Japan Aerospace Exploration Agency, 3-1-1 Yoshinodai, Sagamihara, Kanagawa 229-8510, Japan}


\begin{abstract}
In order to investigate when and how the birth of a protostellar core occurs, 
we made survey observations of four well-studied dense cores in the Taurus molecular cloud using 
CO transitions in submillimeter bands.  
We report here the detection of unexpectedly warm ($\sim$ 30 $-$ 70 K), 
extended (radius of $\sim$ 2400 AU), dense (a few times 10$^{5}$ cm$^{-3}$) gas at the heart of one of 
the dense cores, L1521F (MC27), within the cold dynamically collapsing components.  
We argue that the detected warm, extended, dense gas may originate from shock regions caused 
by collisions between the dynamically collapsing components and outflowing/rotating components 
within the dense core.  We propose a new stage of star formation, "warm-in-cold core stage (WICCS)", 
i.e., the cold collapsing envelope encases the warm extended dense gas at the center due 
to the formation of a protostellar core.  WICCS would constitutes a missing link in evolution between a cold 
quiescent starless core and a young protostar in class 0 stage that has a large-scale 
bipolar outflow. 
\end{abstract}




\section{Introduction}
Low mass stars form through gravitational contraction within cold 
dense 
condensations, called dark cloud dense cores.  
The dense cores remain roughly isothermal as long as radiative cooling is comparable to the 
heating from compression of the medium.  
This compression is due to the gravitational contraction over $\sim$10$^{5}$ years (e.g., Masunaga et al. 1998). 
When the central density reaches 10$^{10-11}$ cm$^{-3}$, compression heating 
becomes dominant and a condensed, more or less star-sized object, 
protostellar core, forms at the heart of a dense core \citep{bos95,mas98}.  

%
A starless dense core is a cold ($\sim$ 10 K) dense ($\ga$ 10$^{4}$ \cc) large ($\sim$ 20,000 AU) condensation 
of gas and dust (e.g., Bergin \& Tafalla 2007). 
It doesn't harbor any embedded infrared point source (hereafter IPS).  
Some evolved starless cores have dynamically collapsing envelopes.  
On the other hand, class 0 objects (e.g., Andr{\'e} et al. 2000), which are embedded in protostellar dense cores with a temperature of $\sim$ 30 K, 
a density of $\ga$ 10$^{5}$ \cc, and a size of $\lesssim$ 10,000 AU, 
are characterized by the existence of IPSs and large-scale molecular bipolar outflows driven by protostars.  
Note that there are dense cores that harbor very low luminosity embedded sources within them but do not have large-scale bipolar outflows.  
These objects were used to be categorized as starless cores, however, embedded sources were revealed 
by deep water maser emission survey at the Nobeyama 45-m telescope 
(e.g., Furuya et al. 2001) 
and deep infrared observations by the Spitzer space telescope 
(e.g., Dunham et al. 2008).  
They are in a stage more evolved than the starless core stage and younger than the class 0 stage.  
The formation of the protostellar cores must occur in such a transition stage.  

In order to investigate when and how the birth of a protostellar core occurs, 
we observed four dense cores in the Taurus molecular cloud using 
two CO lines in submillimeter bands, i.e., $J= 6-5$ 
(wavelength $\lambda$ 433.544 $\mu$m) and $7-6$ ($\lambda$ 371.650 $\mu$m) transitions.  
The energy levels at $J=5$ and $J=6$ 
correspond to 83 K and 116 K, respectively.  The critical densities at the temperatures of 100 K 
traced by the molecule are several times 10$^{5}$ cm$^{-3}$.  
With the molecular lines that trace both warm and dense gas components, 
one can study the physical properties of the regions closely related to 
the final processes of protostellar core formation 
when material at the center starts warming up due to the formation of the protostellar core.  

\section{Source Selection}
Our surveyed 
dense cores include L1521E, L1498, L1521F and L1544.  
These cores are selected because 
(i) they are well studied and known as objects in early stages of star formation, i.e., 
the starless core stage or the stage of a dense core that harbors 
a very low luminosity IPS and doesn't show a large-scale bipolar outflow, and 
(ii) they are reasonably isolated hence good targets to study internal star formation processes 
without any external disturbance.  
Among the aforementioned dense cores, L1521E is in the 
youngest evolutionary stage because of its richness in carbon-chain molecules (e.g., Hirota et al. 2002). All the cores 
except for L1521E show a gas infall signature \citep{hir02,taf98,oni99,taf04}.  
L1498 has a relatively low density of 10$^{4}$ cm$^{-3}$ 
and is in a younger evolutionary stage as compared to L1544 and L1521F.  
L1521F \citep{miz94,cod97,oni99,shi04,cra04} has been 
one of primary targets among dense cores because of the short dynamical time scale due to 
the high density of 10$^{6}$ cm$^{-3}$ at the center \citep{oni99}, chemically evolved features \citep{shi04,cra04}, 
and the discovery of the 
IPS, L1521F-IRS, which may be  a protostar in the making, within the dense core using the Spitzer space telescope \citep{ter05,bou06,ter09}.  
L1544 shows a similar nature to L1521F, however, no IPS has been detected to date.  
Based on the observed properties, L1521F is the most evolved sample in the set.

For L1521F, the Spitzer observation imaged a bipolar nebula of scattered light extending in the east-west direction associated 
with L1521F-IRS \citep{bou06,ter09}, which may trace a cavity.  However, no large-scale bipolar outflow is observed 
using spectroscopy.  A deep centimeter continuum observation confirms that there are no shock-ionized 
inner regions of a bipolar outflow from a $\sim$ 0.1 \Lsun~ source \citep{har02}. 
L1521F-IRS is not categorized as a class 0 source owing to lack of a large-scale bipolar outflow.  
It is in an evolutionary category similar to GF 9-2 (e.g., Furuya et al. 2006, 2009) that shows very weak water maser emission from an embedded source but 
does not have a large-scale bipolar outflow.  

%

\section{Observations}
%
%
The CO observations in submillimeter bands were carried out at the Caltech Submillimeter Observatory (CSO). 
The CO $J= 6-5$ (frequency $\nu$ 691.473076GHz, Goldsmith et al. 1981) and 
$7-6$  ($\nu$ 806.651801GHz, Schultz et al. 1985) 
data were taken on 2006 February 06 $-$ 08 and on 2009 January 22 UT, respectively.  
The high-altitude dry site, sensitive receivers and high efficiency of the telescope combined to permit the observations. 
The Dish Surface Optimization System was used during the observations to compensate for the gravitational 
deformation of the 10.4-m diameter reflector (Leong et al. 2005). 
The telescope's pointing was checked every one -- two hours  using planets.  
The pointing accuracy is estimated to be $\sim$ 3\arcsec~ for both observations and  
the accuracy of the map registration of the two CO transitions is within 3\arcsec.  

Cryogenically cooled SIS receivers operation at 4 K at the CSO that we used for the observations 
produced typical single sideband 
system temperature of $\sim$ 3900 K at 433.5 $\mu$m and $\sim$ 2300 -- 3000 K at 371.7 $\mu$m \citep{koo00} 
measured with a 50 MHz bandwidth 
spectrometer. Beam chopping method was used. 
The beam sizes at the two wavelengths are about 10\arcsec and 9\arcsec at 433.5 and 371.7 $\mu$m, respectively. 
From observations of Mars and Saturn, the main 
beam efficiencies are measured to be $\sim$30 \% and 
$\sim$35 \% at 433.5 $\mu$m and 371.7 $\mu$m at elevations of 30 and 48 degrees, respectively.  
The RMS noise levels for both observations are about 0.6 -- 1.4 K in \tmb.  
The velocity resolution of both observations is $\sim$ 0.02 \kms.




\section{Results}\label{results}
\begin{figure}
\epsscale{.70}
\includegraphics[angle=0,scale=1.3]{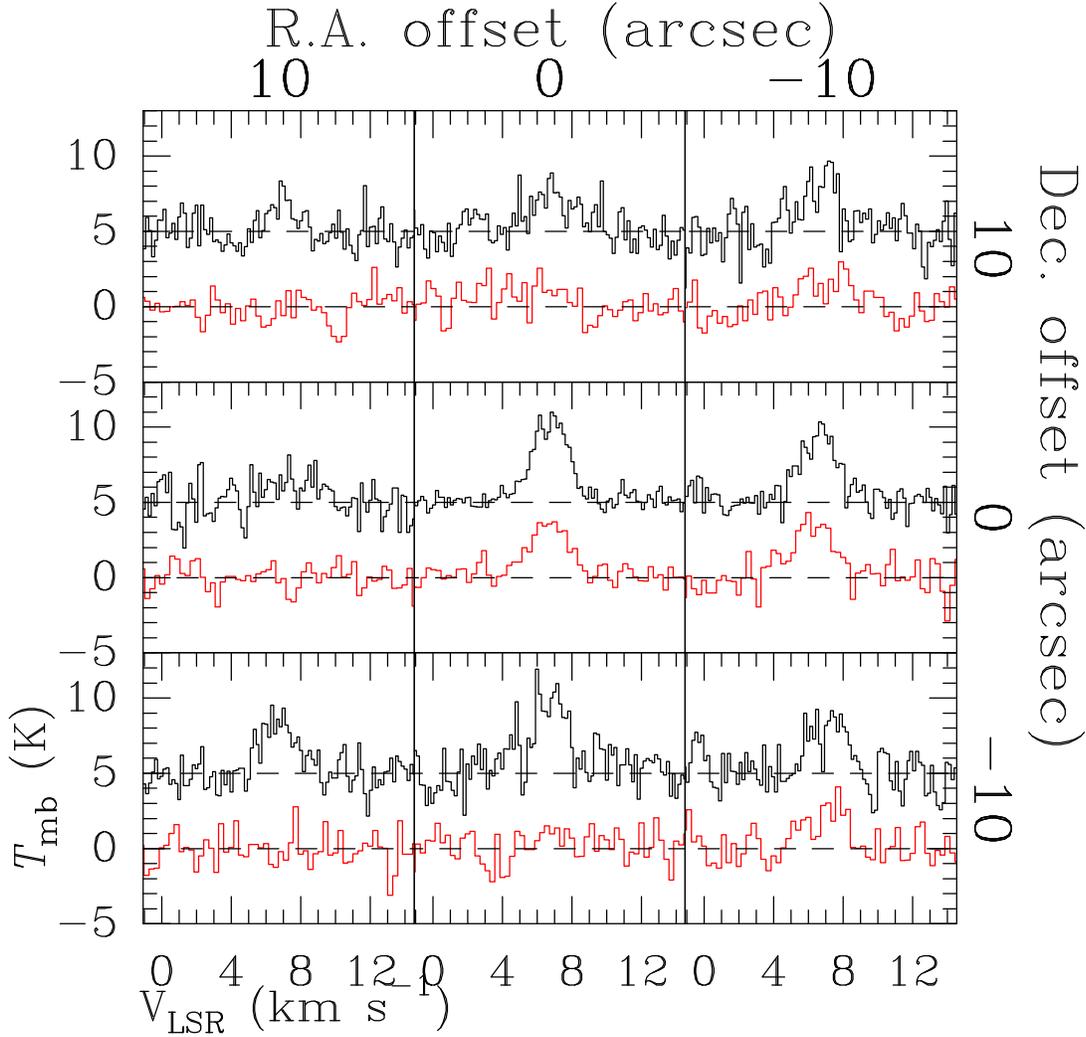} 
\caption{Profile maps of the CO $6-5$ (black line) and $7-6$ (red line) emission of L1521F. 
The central panel (i.e., the origin of the diagram) corresponds to 
the position at R.A. = 4h 25m 34.s29, Dec. = 26\deg 45\arcmin 8.7\arcsec (B1950.0), where the peak intensities of the two transitions are observed.
The peak \Tmb~ values of the $6-5$ and $7-6$ emission are measured to be 5.5 and 3.7 K with observed line widths \VFWHM~ of 2.3 and 2.6 \kms~ by Gaussian fitting, respectively. 
Note that L1521F-IRS is located near the position of (R.A. offset, Dec. offset) = (0\arcsec, -10\arcsec).  
 \label{co76profilemap}}
\end{figure}

Among the observed dense cores, only one, L1521F, shows significant CO line emission 
at both transitions (Figure
\ref{co76profilemap}). 
The rest of the dense cores don't show the CO $6-5$ line emission brighter than 0.3 K in the observed main beam temperatures (\tmb).  
Observed properties of CO lines of L1521F are summarized in Table \ref{tbl-1}.  
\Tmb~ values of the CO transitions are weak (see Table \ref{tbl-1}), 
suggesting that the emission may be optically thin.  
The peak \Tmb~ is observed at the position of 
R.A. = 4$^{\rm h}$ 25$^{\rm m}$ 34.$^{\rm s}$29, Dec. = 26\deg 45\arcmin 8\arcsec.7 (B1950.0) for both transitions.  
Note that 
L1521F-IRS is located at 9" south of the peak position.
Considering the thermal velocity width of the CO gas at 100 K is only about 0.5 \kms, 
a significant fraction of the broad velocity widths originates from non-thermal motions.

We estimate the temperature of the observed warm dense gas components. 
Taking advantage of the similar beam sizes of the observations of the two transitions, one can 
derive the excitation temperature ($T_{\rm ex}$) using the following equation (e.g., Shinnaga et al. 2008): 
\begin{equation}
T_{\rm ex} = \frac{38.714~{\rm K}}{\ln\left[\left(\frac{7}{6}\right)^2\frac{T_{\rm{mb~ CO}~ 6-5}}{T_{{\rm mb~ CO}~ 7-6}}\right]}  ~~~~. 
\end{equation}
\Tex~ values estimated at the positions 
of (R.A. offset, Dec. offset) = (0\arcsec, 0\arcsec), (-10\arcsec, +10\arcsec), (-10\arcsec, 0\arcsec), and (-10\arcsec, -10\arcsec)
are 57 $\pm$ 1.7, 40 $\pm$ 7, 74 $\pm$ 19, and 68 $\pm$ 5.5 K, respectively. 
Note that the CO $7-6$ emission is not detected towards the position of L1521F-IRS, 
indicating that the temperature near the IPS is still low ($\lesssim$ 30 K).  
Over the central 
30\arcsec $\times$ 30\arcsec~ region (corresponds to 4200 AU), the averaged $T_{\rm ex}$ is calculated to be 34.4 $\pm$ 0.94 K.  
These temperatures are much higher than the \Tex~ measured with an \n2hp transition, 5.0 K \citep{shi04}. 

\begin{figure}
\epsscale{.70}
\includegraphics[angle=-90,scale=0.7]{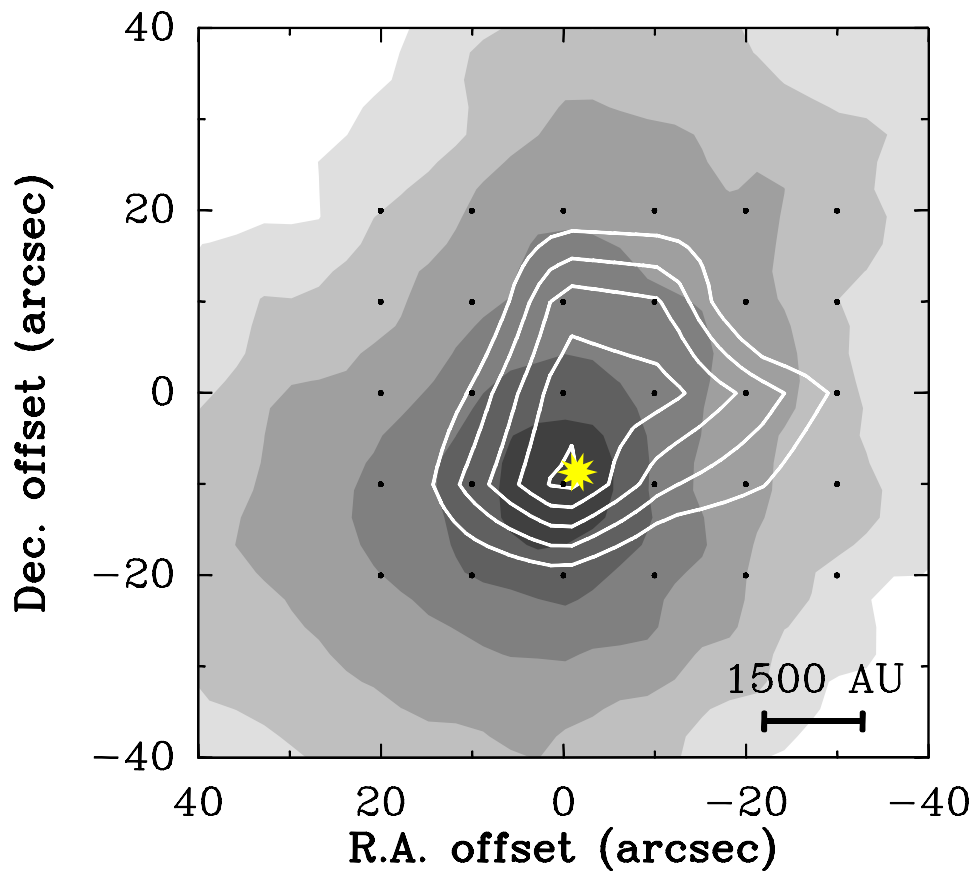} 
\caption{
The total integrated intensity map of CO $6-5$ emission that traces WEDG 
(white contours), overlaid on the grey-scale contour map of the 
cold extended components traced by the dust continuum emission at 850 $\mu$m  \citep{shi04}.  
The contours of WEDG are drawn at 5, 7, 9, 11 and 13 $\sigma$ levels (1 $\sigma$ corresponds to 1.1 K \kms). 
The map origin is the same as that of Figure \ref{co76profilemap}. The yellow star marks the location of L1521F-IRS.   
The black dots represent the observed positions. The beam sizes of the CO $6-5$ and 850 $\mu$m 
continuum maps are 10\arcsec~ and 14\arcsec, respectively.
\label{co65map}}
\end{figure}

Figure \ref{co65map} presents the total integrated intensity map of the CO $J= 6-5$ emission 
that traces the warm extended dense gas (hereafter WEDG), overlaid on 
the dust continuum map at 850 $\mu$m  that traces the cold ($\sim$10 K) extended (16,000 AU) 
dense ($\ga$ 10$^{5}$ cm$^{-3}$) condensation \citep{shi04}. 
The observed effective radius of WEDG within the 3 $\sigma$ contour is 2400 AU (0.012 pc).  
WEDG sits in the central region of the cold condensation and is distributed somewhat asymmetrically, 
extending to the north and northwest directions from L1521F-IRS.  The peak \Tmb~  
of the CO $6-5$ 
transition is found at 
the position of (R.A. offset,  Dec. offset) = (0\arcsec, 0\arcsec),  
while the CO $6-5$ total integrated intensity map 
peaks at the position near L1521F-IRS, i.e., (R.A. offset,  Dec. offset) = (0\arcsec, -10\arcsec) 
owing to a large line width measured at the position (observed line width \VFWHM~ of 2.46 $\pm$ 0.46 \kms). 

%
%
%

\section{Discussion}\label{discussion}
We considered four candidate mechanisms for the excitation of WEDG.   These are radiative excitation from L1521F-IRS, 
infrared pumping from L1521F-IRS, heating by UV radiation generated from a possible accretion 
disk, like the one that may explain narrow submillimeter CO line components associated with class 0 objects \citep{spa95}, and 
heating due to shocks.  Only the last one provides a convincing 
interpretation of the data.  Radiative excitation from L1521F-IRS is dismissed as L1521F-IRS's luminosity 
is not high enough to pump CO molecules up to such a high $J$ level over $\ga$ 2400 AU region.  
In fact, CO $7-6$ 
emission is not detected near the L1521F-IRS position.  Infrared pumping is ruled out due to the large vibrational 
level spacing of CO ($\sim$3000 K, Carroll et al. 1981). The protostellar core's temperature doesn't become higher than 3000 K 
as molecular hydrogen begins to dissociate at $\sim$ 2000 K.  
It is implausible that UV radiation from 
the accretion disk heats the observed CO gas 
as the peak \Tmb~ values of both transitions are not found towards the position of the IPS.  

On the other hand, shocks would be able to pump the gas up to $\sim$ 100 K over a large region \citep{neu95}. 
In fact, a systematic velocity pattern is not measured in WEDG, indicating that WEDG may be excited 
in shock-heated regions generated by the collision between the cold dynamically collapsing components 
and outflowing/rotating components near the center.
Considering the presence of L1521F-IRS, it's highly likely 
that some outflow activities have already been initiated.  The cavity of the bipolar nebula also indicates the 
existence of a small jet associated with the protostar.  
Furthermore, L1521F-IRS may have a large rotating 
circumstellar disk with a size of the order of 10$^{3}$ AU.  The north-south elongated feature seen in the 850 $\mu$m 
dust continuum map may indicate the existence of the north-south elongated disk that is perpendicular to the 
cavity axis, i.e., perpendicular to the axis of the bipolar nebula.  
Based on the measured velocity gradient of 15 km s$^{-1}$ pc$^{-1}$ over  
0.01 pc with a clump mass of 0.1 \Msun~ by using the \n2hp $J=1-0$ transition at the center of the dense core \citep{shi04},
the centrifugal radius of the collapsing medium is estimated to be about 10$^{3}$ AU.   
In addition, the asymmetric distribution of WEDG supports the view that WEDG comes from the shocks but 
not from the small jet itself.  
If the two CO lines are in the LTE condition, the lower limit of the mass of WEDG becomes an order of 10$^{-2}$ \Msun.  

\begin{figure}
\plotone{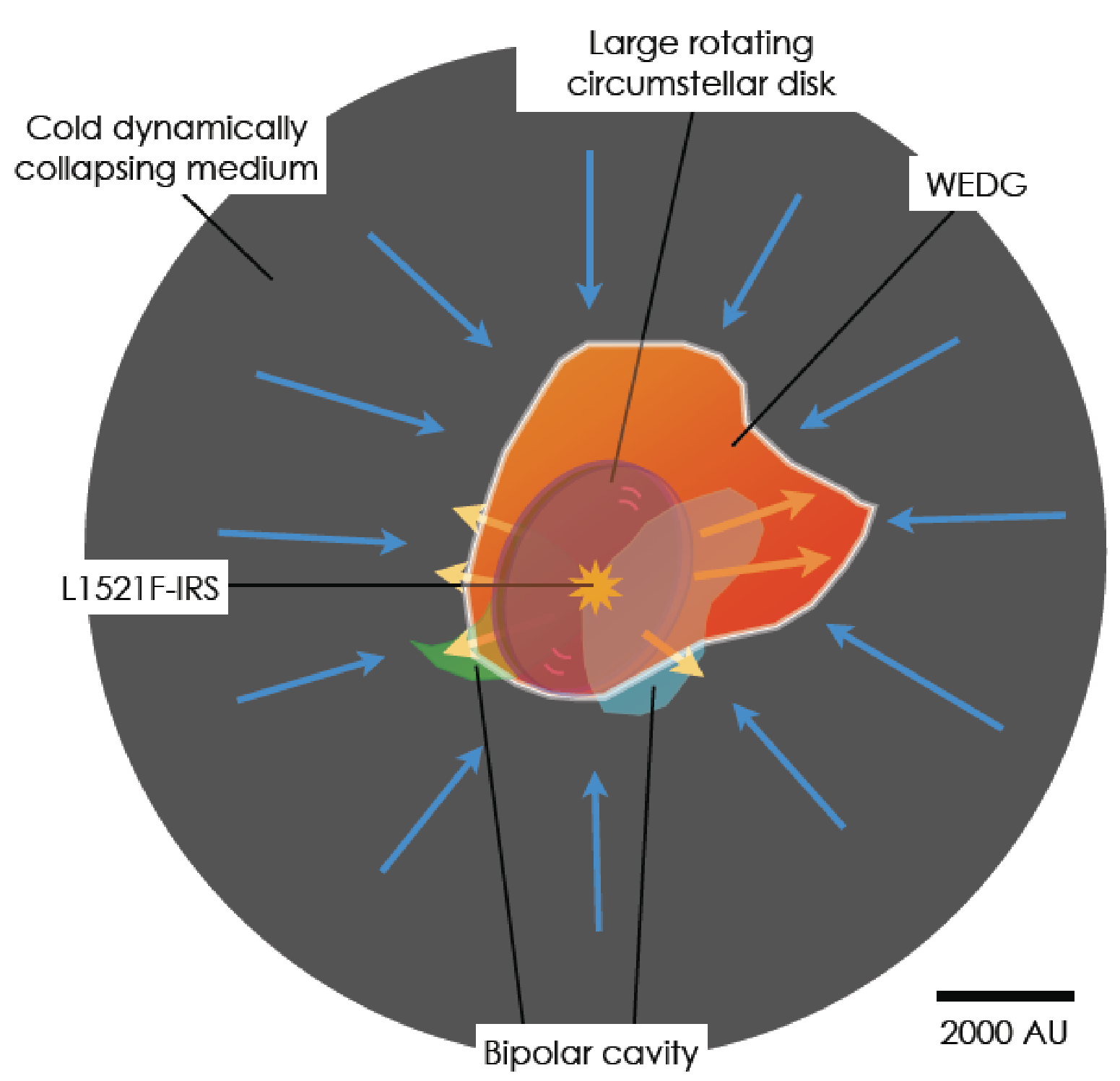} 
\caption{
Schematic illustration of the heart of L1521F.  
The blue arrows indicate the cold dynamically infalling medium.  
The white contour filled with red color represents the lowest contour of WEDG shown in Figure \ref{co65map}.  
The yellow star marks the location of L1521F-IRS. The irregular shaped objects with green and blue colors associated with L1521F-IRS indicate the bipolar cavity. The large rotating disk is perpendicular to the cavity axis. The orange arrows pointing outwards through the cavity indicate 
the outflowing gas from L1521F-IRS.  
\label{cartoon}}
\end{figure}

Figure \ref{cartoon} illustrates a magnified view of the center of the dense core that we propose. 
The cavity with blue color may be in front of the disk, while the cavity with green color is behind the disk, 
along the line of sight.  The cavity in front is opening towards the west from L1521F-IRS, which may channel part 
of the outflowing gas, making the shock easier to observe on the western side.  
The north-south extension of WEDG 
may be due to the circumstellar disk elongated along the north-south direction.


\section{Summary and Future Work} \label{summary}
We made survey observations using the two CO transitions in submillimeter bands  
to search for 
the dark cloud dense cores which harbor newly born protostellar cores 
and to investigate the physical properties of the objects.  
This study identified a transient stage of star formation between the starless core and class 0 stages.  
As the starless core stage progresses, 
the cold collapsing dense material starts warming up at the center due to the formation of a protostellar core 
and forms WEDG in the central region.  
We name this new stage "warm-in-cold core stage (WICCS)".  
One should search for 
WEDG using warm and dense gas tracers such as the CO $6-5$ and $7-6$ transitions to identify the objects in WICCS.  
%
This object would constitute a missing link in evolution between a starless core 
and a protostar, yielding an important step toward understanding of the formation 
mechanism of a protostellar core and a protostar.  
Survey observations of CO transitions in submillimeter bands towards  
dense cores in early evolutionary stages are necessary to add more samples in WICCS 
in order to obtain a complete picture of WICCS.  




\acknowledgments
This research was performed at the Caltech Submillimeter Observatory, supported by NSF grant 
AST-05-40882 and AST-0838261. 
This work was also supported by Grant-in-Aids from the Ministry of
Education, Culture, Sports, Science and Technology of Japan (No.
19204020 and No. 20740113).
H. S. is grateful to Richard Chamberlin, Brian Force, and Hiroshige Yoshida for their support on the observations 
in 2009 January.

\clearpage
\begin{deluxetable}{ccccc}
\tabletypesize{\scriptsize}
\tablecaption{Observed Properties of the CO Lines \label{tbl-1}}
\tablewidth{0pt}
\tablehead{
\colhead{Transition} & \colhead{Wavelength} & \colhead{\Tmb \tablenotemark{a} } & \colhead{$V_{\rm FWHM}$\tablenotemark{b}}   & \colhead{$V_{\rm LSR}$\tablenotemark{c}}   \\%
\colhead{}     & \colhead{($\mu$m)} & \colhead{(K)} & \colhead{(\kms)}   & \colhead{(\kms)} 
}
\startdata
$J=6-5$ & 433.544 & 1.5 -- 5.5 & 1.1 -- 4.2  & 6.77 (3)   \\
$J=7-6$ & 371.650 & 1.6 -- 3.7 & 2.5 -- 3.5  & 6.66 (8) \\
%
 \enddata
 
\tablenotetext{a}{Observed peak main beam temperatures. }
\tablenotetext{b}{Observed line widths.}
\tablenotetext{c}{Intensity-weighted mean velocities at the peak intensity position. The numbers in parentheses represent 3 $\sigma$ 
in units of 0.01 \kms.}  
\end{deluxetable}
\clearpage



\end{document}